\documentclass[journal,comsoc,10pt]{IEEEtran}

\usepackage{cite}
\usepackage[english]{babel}
\usepackage{blindtext}
\usepackage{tikz,pgfplots}
\usepackage{tkz-euclide}
\pgfplotsset{compat=1.5}
\usetikzlibrary{calc}
\usepgfplotslibrary{polar}

\usepackage{epstopdf} 
\usepackage{amsmath}
\usepackage{amsfonts}
\usepackage{wrapfig}
\usepackage{mathrsfs}
\usepackage{accents}
\usepackage{acronym}
\usepackage{graphicx}
\usepackage{textcomp}
\usepackage{xcolor}
\usepackage{mathtools}
\usepackage{color,soul, colortbl}
\usepackage{graphicx}
\usepackage{booktabs}
\usepackage{multirow}
\usepackage[font={footnotesize}]{caption}
\usepackage[font={footnotesize}]{subcaption}
\usepackage{algorithm}
\usepackage{algorithmic}

\usepackage{caption}
\usepackage{subcaption}
\usepackage{url}

\newcommand{\fig}[1]{Fig.~\ref{#1}}
\newcommand{\tab}[1]{Tab.~\ref{#1}}
\newcommand{\secref}[1]{Section~\ref{#1}}

\acrodef{prop}[\textit{MIMORPH}]{MIMO Radio Platform for Heterogeneous wireless systems}

\acrodef{abft}[A-BFT]{Association Beamforming Training}
\acrodef{ack}[ACK]{Acknowledge}
\acrodef{adc}[ADC]{Analog-to-Digital Converter}
\acrodef{aoa}[AoA]{Angle of Arrival}
\acrodef{aod}[AoD]{Angle of Departure}
\acrodef{ap}[AP]{Access Point}
\acrodef{amc}[AMC]{Advanced Mezzanine Card}
\acrodef{awv}[AWV]{Antenna Wave Vector}
\acrodef{axi}[AXI]{Advanced eXtensible Interface}
\acrodef{ber}[BER]{Bit Error Rate}
\acrodef{bft}[BFT]{Beamforming Training}
\acrodef{bp}[BP]{Beam Pattern}
\acrodef{brp}[BRP]{Beam Refinement Phase}
\acrodef{cs}[CS]{Compressed Sensing}
\acrodef{cdf}[CDF]{Cumulative Distribution Function}
\acrodef{cef}[CEF]{Channel Estimation Field}
\acrodef{cfo}[CFO]{Carrier Frequency Offset}
\acrodef{cir}[CIR]{Channel Impulse Response}
\acrodef{csi}[CSI]{Channel State Information}
\acrodef{csirs}[CSI-RS]{CSI-Reference Signal}
\acrodef{cv}[CV]{Constant Velocity}
\acrodef{cvi}[CV]{computer vision}
\acrodef{cnn}[CNN]{Convolutional Neural Network}
\acrodef{cots}[COTS]{Commercial-Off-The-Shelf}
\acrodef{dft}[DFT]{Discrete Fourier Transform}
\acrodef{dl}[DL]{Deep Learning}
\acrodef{dma}[DMA]{Direct Memory Access}
\acrodef{dmg}[DMG]{Directional Multi Gigabit}
\acrodef{dti}[DTI]{Data Transfer Interval}
\acrodef{edmg}[EDMG]{Enhanced Directional Multi Gigabit}
\acrodef{ekf}[EKF]{Extended Kalman Filter}
\acrodef{fpga}[FPGA]{Field Programmable Gate Array}
\acrodef{fmcw}[FMCW]{Frequency-Modulated Continuous-Wave}
\acrodef{fov}[FoV]{Field-of-View}
\acrodef{ft}[FT]{Fourier Transform}
\acrodef{gpio}[GPIO]{General Purpose Input/Output}
\acrodef{gsps}[GSPS]{Giga-Samples per Second}
\acrodef{har}[HAR]{Human Activity Recognition}
\acrodef{ht}[HT]{High Throughput}
\acrodef{if}[IF]{Intermediate Frequency}
\acrodef{ifs}[IFS]{Inter-Frame Spacing}
\acrodef{iht}[IHT]{Iterative Hard Thresholding}
\acrodef{isac}[ISAC]{Integrated Sensing And Communication}
\acrodef{jcs}[JCS]{Joint Communication \& Sensing}
\acrodef{jpdaf}[JPDAF]{Joint Probabilistic Data Association Filter}
\acrodef{los}[LOS]{Line-of-Sight}
\acrodef{lbm}[LBM]{Loop-Back Memory}
\acrodef{mae}[MAE]{Mean Absolute Error}
\acrodef{mcs}[MCS]{Modulation and Coding Scheme}
\acrodef{md}[$\mu$D]{micro-Doppler}
\acrodef{mimo}[MIMO]{Multiple Input Multiple Output}
\acrodef{ml}[ML]{Machine Learning}
\acrodef{mmwave}[mmWave]{Millimeter-Wave}
\acrodef{msps}[MSPS]{Mega-Samples per Second}
\acrodef{mu}[MU]{Multiple User}
\acrodef{MUSIC}[MUSIC]{MUlti SIgnal Classification}
\acrodef{nac}[NAC]{Normalized Auto Correlation}
\acrodef{nco}[NCO]{Numerical Controlled Oscillator}
\acrodef{nlos}[NLOS]{Non-Line-of-Sight}
\acrodef{ofdm}[OFDM]{Orthogonal Frequency Division Multiplexing}
\acrodef{per}[PER]{Packet Error Rate}
\acrodef{phy}[PHY]{Physical Layer}
\acrodef{pl}[PL]{Programmable Logic}
\acrodef{pov}[POV]{Point-of-View}
\acrodef{ps}[PS]{Processing System}
\acrodef{rf}[RF]{Radio Frequency}
\acrodef{rfsoc}[RFSoC]{Radio Frequency System on a Chip}
\acrodef{rss}[RSS]{Received Signal Strength}
\acrodef{rom}[ROM]{Read Only Memories}
\acrodef{sc}[SC]{Single Carrier}
\acrodef{sdr}[SDR]{Software Defined Radio}
\acrodef{siso}[SISO]{Single Input Single Output}
\acrodef{sls}[SLS]{Sector Level Sweep}
\acrodef{snr}[SNR]{Signal-to-Noise Ratio}
\acrodef{soc}[SoC]{System on a Chip}
\acrodef{spb}[SPB]{Signal Processing Blocks}
\acrodef{srrc}[SRRC]{Square-Root-Raised-Cosine}
\acrodef{ssb}[SSB]{Synchronization Signal Block}
\acrodef{ssr}[SSR]{Super Sample Rate}
\acrodef{sta}[STA]{Station}
\acrodef{stf}[STF]{Short Training Field}
\acrodef{stft}[STFT]{Short Time Fourier Transform}
\acrodef{su}[SU]{Single User}
\acrodef{tf}[TF]{Time-Frequency}
\acrodef{toa}[ToA]{Time of Arrival}
\acrodef{usrp}[USRP]{Universal Software Radio Peripheral}
\acrodef{vht}[VHT]{Very High Throughput}
\acrodef{wlan}[WLAN]{Wireless Local Area Network}

\newcommand{\rev}[1]{{\color{blue}#1}}
\renewcommand{\rev}[1]{#1}

\usepackage{scalerel}
\usepackage{tikz}
\usetikzlibrary{svg.path}

\definecolor{orcidlogocol}{HTML}{A6CE39}
\tikzset{
  orcidlogo/.pic={
    \fill[orcidlogocol] svg{M256,128c0,70.7-57.3,128-128,128C57.3,256,0,198.7,0,128C0,57.3,57.3,0,128,0C198.7,0,256,57.3,256,128z};
    \fill[white] svg{M86.3,186.2H70.9V79.1h15.4v48.4V186.2z}
                 svg{M108.9,79.1h41.6c39.6,0,57,28.3,57,53.6c0,27.5-21.5,53.6-56.8,53.6h-41.8V79.1z M124.3,172.4h24.5c34.9,0,42.9-26.5,42.9-39.7c0-21.5-13.7-39.7-43.7-39.7h-23.7V172.4z}
                 svg{M88.7,56.8c0,5.5-4.5,10.1-10.1,10.1c-5.6,0-10.1-4.6-10.1-10.1c0-5.6,4.5-10.1,10.1-10.1C84.2,46.7,88.7,51.3,88.7,56.8z};
  }
}

\newcommand\orcidicon[1]{\href{https://orcid.org/#1}{\mbox{\scalerel*{
\begin{tikzpicture}[yscale=-1,transform shape]
\pic{orcidlogo};
\end{tikzpicture}
}{|}}}}

\usepackage[hidelinks]{hyperref} 
\usepackage{cleveref}

\begin{document}

\title{DISC: a Dataset for Integrated Sensing and Communications in mmWave Systems
\thanks{$^{\dag}$~These authors are with the Department of Information Engineering, University of Padova, Italy (email: \texttt{jacopo.pegoraro@unipd.it}).  
\newline $^*$~These authors are with the IMDEA Networks Institute, Madrid, Spain.
}
}
\author{
Jacopo Pegoraro$^{\dag}$\textsuperscript{\orcidicon{0000-0003-3555-5666}}, Pablo Saucedo$^*$\textsuperscript{\orcidicon{0009-0003-1010-732X}},
Jesus O. Lacruz$^*$\textsuperscript{\orcidicon{0000-0002-6641-2003}}, \\
Michele Rossi$^\dag$\textsuperscript{\orcidicon{0000-0003-1121-324X}}, Joerg Widmer$^*$\textsuperscript{\orcidicon{0000-0001-6667-8779}} 
\vspace{-0.5cm}}

\hyphenation{op-tical net-works semi-conduc-tor}

\maketitle

\begin{abstract}
This paper presents DISC, a dataset of millimeter-wave channel impulse response measurements for integrated human activity sensing and communication. This is the first dataset collected with a software-defined radio testbed that transmits 60~GHz IEEE~802-11ay-compliant packets and estimates the channel response including scattered signals off the moving body parts of subjects moving in an indoor environment. \rev{The provided data consists of three parts, for more than 2 hours of channel measurements with high temporal resolution (0.27~ms inter-packet time)}. DISC contains the contribution of 7 subjects performing 5 different activities, and includes data collected from two distinct environments. Unlike available radar-based millimeter-wave sensing datasets, our measurements are collected using uniform packet transmission times and sparse traffic patterns from real \mbox{Wi-Fi} deployments. We develop, train, and release open-source baseline algorithms based on DISC to perform human sensing tasks. Our results demonstrate that DISC can serve as a multi-purpose benchmarking tool for machine learning-based human activity recognition, radio frequency gait analysis, and sparse sensing algorithms for next-generation integrated sensing and communications.
\end{abstract}


\IEEEpeerreviewmaketitle

\section{Introduction}\label{sec:intro}

\IEEEPARstart{D}{espite} the huge interest towards \ac{isac} systems working in the \ac{mmwave} frequency band, there is a lack of public datasets in this regard. Thus, it is hard for researchers to develop and \textit{validate} signal processing or \ac{ml}-based \ac{isac} algorithms beyond simulation environments.
Unlike widely studied Wi-Fi sensing in the \mbox{sub-$6$~GHz} band, \ac{mmwave} \ac{isac} lacks affordable commercial devices, tools for the extraction of \ac{csi}~\cite{gringoli2019free}, so it has mostly relied on simulation tools~\cite{blandino2022tools}. Indeed, most of the experimental works in the field use \ac{mmwave} radar devices which employ specifically optimized waveforms for each application and are inapt for communications \cite{singh2019radhar}. Moreover, each work typically uses a different dataset and baseline algorithms. With the rising importance of \ac{dl} for radio signal processing, it becomes key to foster easier comparison to state-of-the-art algorithms on common data. This is especially true since physical-layer radio signals are cumbersome to collect and store, due to the high sampling rates and data volume. 

\textit{Challenges.} In light of the above discussion, we identify two main unaddressed challenges we solve with the dataset and algorithms presented in this paper. First, there is a lack of datasets that contain radio signal traces to be used for the design and validation of \ac{mmwave} \ac{isac} algorithms. Key aspects are the size and diversity of such datasets, as most wireless sensing applications require data-hungry \ac{dl} models to extract complex features from the reflected signal while generalizing to different environments and sensing subjects. Moreover, the data traces should present the typical characteristics of \ac{isac} systems that can not be found in radar datasets, such as \textit{(i)} diverse, irregular, and sparse traffic patterns, and \textit{(ii)} the usage of standard-compliant communication waveforms. Existing efforts in this sense are limited to simulation tools that lack the complexity and hardware impairments of real measurements~\cite{blandino2022tools}.

Second, open-source benchmark signal processing and \ac{dl} algorithms to perform sensing applications are needed. These should encompass the variety of \ac{isac} applications, such as people localization and tracking, \ac{har}, and gait identification. Moreover, challenging sensing scenarios characterized by resource-constrained channel acquisition and diverse multipath environments should be investigated, building on public baseline algorithms.
Solving these challenges is a key enabler for \ac{isac} research, which has been identified as a core feature of next-generation $6$G mobile networks~\cite{yang2024integrated} and \acp{wlan}~\cite{chen2023wifi}.

\textit{DISC.} \rev{In this paper we present DISC \cite{disc}, a \ac{mmwave} \ac{isac} dataset containing \ac{cir} measurements from standard-compliant \ac{sc} IEEE~802.11ay packets.} The \ac{cir} sequences contain backscattered copies of the transmitted packets on people moving in an indoor environment.
We solve both open challenges by \textit{(i)}~providing a large-scale dataset of realistic channel estimates to perform human sensing tasks, called DISC, containing $60$~GHz IEEE~802.11ay Wi-Fi channel estimates including different people, environments, and traffic patterns~\cite{disc}, and \textit{(ii)}~making a set of benchmark algorithms for people tracking, human activity recognition, and sparse \ac{md} spectrum estimation available to the community, open-sourcing their code implementation. 

The dataset consists of three parts. The first one, DISC-A, contains more than $1$ hour of IEEE~802.11ay \ac{cir} sequences including backscattered signals from $7$ subjects performing $4$ different activities in front of the \ac{isac} transceiver. This part is characterized by uniform packet transmission times, with a granularity of over $3$ \ac{cir} estimates per millisecond, yielding extremely high temporal resolution.
The second part, \mbox{DISC-B}, contains $40$ minutes of \ac{cir} sequences obtained at uniform packet transmission times, with $1$ to $5$ subjects concurrently, freely moving in the environment and performing $5$ different activities. Moreover, in this second part, we use the directional transmission capabilities of our testbed to allow the estimation of the \ac{aoa} of the backscattered signal, which enables \textit{tracking} the subject across time. DISC-B also includes data collected in a different test environment, to offer the opportunity to test the generalization capabilities of \ac{dl}-based sensing algorithms.
In the third part, \mbox{DISC-C}, we use open-source data on Wi-Fi traffic patterns to tune the inter-packet duration and collect more realistic \textit{sparse} \ac{cir} sequences. The resulting \ac{cir} measurements, for a total of $9$~minutes, are collected with a single subject performing the same $4$ activities included in the first part.

\textit{Usage.} We envision our dataset and algorithms being used to train and validate new fine-grained sensing approaches. Possible use cases include, but are not limited to, tracking of multiple subjects, extraction of the \ac{md} signatures of human movement from the \ac{cir} \cite{chen2006micro}, which enables \ac{dl}-based \ac{har} \cite{singh2019radhar}, and person identification from individual gait features \cite{vandersmissen2018indoor}. In addition, advanced \ac{isac} problems such as the sparse reconstruction of sensing parameters from irregularly sampled signal traces, domain adaptation from regularly sampled signals to sparse ones, and target tracking under missing measurements can also be tackled using DISC~\cite{pegoraro2022sparcs}. 
\rev{Moreover, DISC is aligned with ISAC standardization efforts by the 3GPP, IEEE~802.11bf, and ETSI, which have shown interest in modeling human-related propagation characteristics of the channel~\cite{yang2024integrated}. In this sense, the availability of experimental data involving reflections on human subjects can support the channel modeling effort.}

The paper is organized as follows. In \secref{sec:11ay} we discuss the necessary preliminaries regarding the IEEE~802.11ay \ac{cir}, while in \secref{sec:testbed} we present the experimental setup including the \ac{isac} \ac{sdr} testbed and the parameters of the experiments.
\secref{sec:dataset} contains an overview of the dataset and in \secref{sec:discussion} we discuss possible research problems that can be addressed by using it. Finally, in \secref{sec:benchmarks} we present results obtained with the proposed benchmark algorithms on DISC. Concluding remarks are given in \secref{sec:conclusion}.

\section{Experimental testbed decription}\label{sec:testbed}

In this section, we describe the experimental setup used to collect the dataset, including the \ac{isac} testbed, the system parameters, and the collection environment.

\subsection{Physical layer system operation and parameters}\label{sec:11ay}

\rev{Our dataset contains \ac{cir} measurements collected according to the IEEE 802.11ay standard \cite{802.11ay}, using a \ac{sc} waveform in single-input single-output mode}. The \ac{cir} contains the complex channel gains for different delays. The delay resolution of the system is given by $\Delta \tau = 1/B$ with $B=1.76$~GHz being the bandwidth of the transmitted signal, spanning an IEEE~802.11ay channel. As a result, the delay resolution of the system is given by $\Delta \tau = 1 / B= 0.568$~ns.
In IEEE 802.11ay the communication is highly \textit{directional}, thanks to the use of phased antenna arrays with narrow \acp{bp}. In addition, \textit{in-packet} beam tracking is supported \cite{802.11ay}, which allows switching the \ac{bp} within a single packet. The different \acp{bp} used illuminate targets located in the environment differently, thus introducing diversity that can be exploited to compute the \ac{aoa} of the propagation paths with high accuracy, see, e.g., \cite{pegoraro2023rapid}. From the beamwidth of the \acp{bp}, one can obtain the angular resolution of the system, $\Delta \theta$, which is approximately~$8^{\circ}$. Beam tracking is done by appending a specific field of pilot symbols, called TRN field, to the packet.  
A TRN field is composed of a tunable number of TRN units, each formed by $6$ complementary Golay sequences of type a (``Ga'') and b (``Gb'') Golay sequences of $128~\pi/2$~BPSK modulated samples, for a total of $768$~samples \cite{802.11ay}. \rev{At the receiver, the signal corresponding to each TRN unit is correlated with the known pilot sequence to estimate the \ac{cir}~\cite{liu2013digital}}.
\rev{Note that exploiting TRN fields with different \acp{bp} for sensing is also featured in IEEE~802.11bf, so DISC can be used to test its physical layer sensing capabilities.}

We denote by $T$ the (tunable) \ac{ifs}, i.e., the time interval between two subsequent channel estimation instants using pilot signals from different packets.
For reliable \ac{md} extraction using the \ac{stft} without aliasing \cite{pegoraro2023rapid}, it is advisable to set $T$ to a value that allows capturing the range of velocities typically covered by human movement in indoor environments. These can reach up to $\pm 5$~m/s for running or other fast movements \cite{vandersmissen2018indoor}. 
In our dataset, when the \ac{ifs} is constant, we set it to $T = 0.27$~ms, which is suitable to capture $v_{\rm max}= \pm c/(4f_o T)\approx \pm 4.48$~m/s. When using a \ac{stft} window $W=64$, one obtains a velocity resolution of $\Delta v= c/(2f_o W T)\approx 0.14$~m/s. In the sequences with irregular inter-packet time, instead, this is determined by realistic traffic patterns of real Wi-Fi access points, as described in \secref{sec:dataset}.
We use periodically transmitted in-packet beam tracking frames with a variable number of TRN units, depending on the part of the dataset considered, and antenna beams covering a \ac{fov} of $[-45^\circ, 45^\circ]$.

\subsection{Data collection environments}\label{sec:env}
\begin{figure}
     \centering
     \includegraphics[width=0.95\columnwidth]{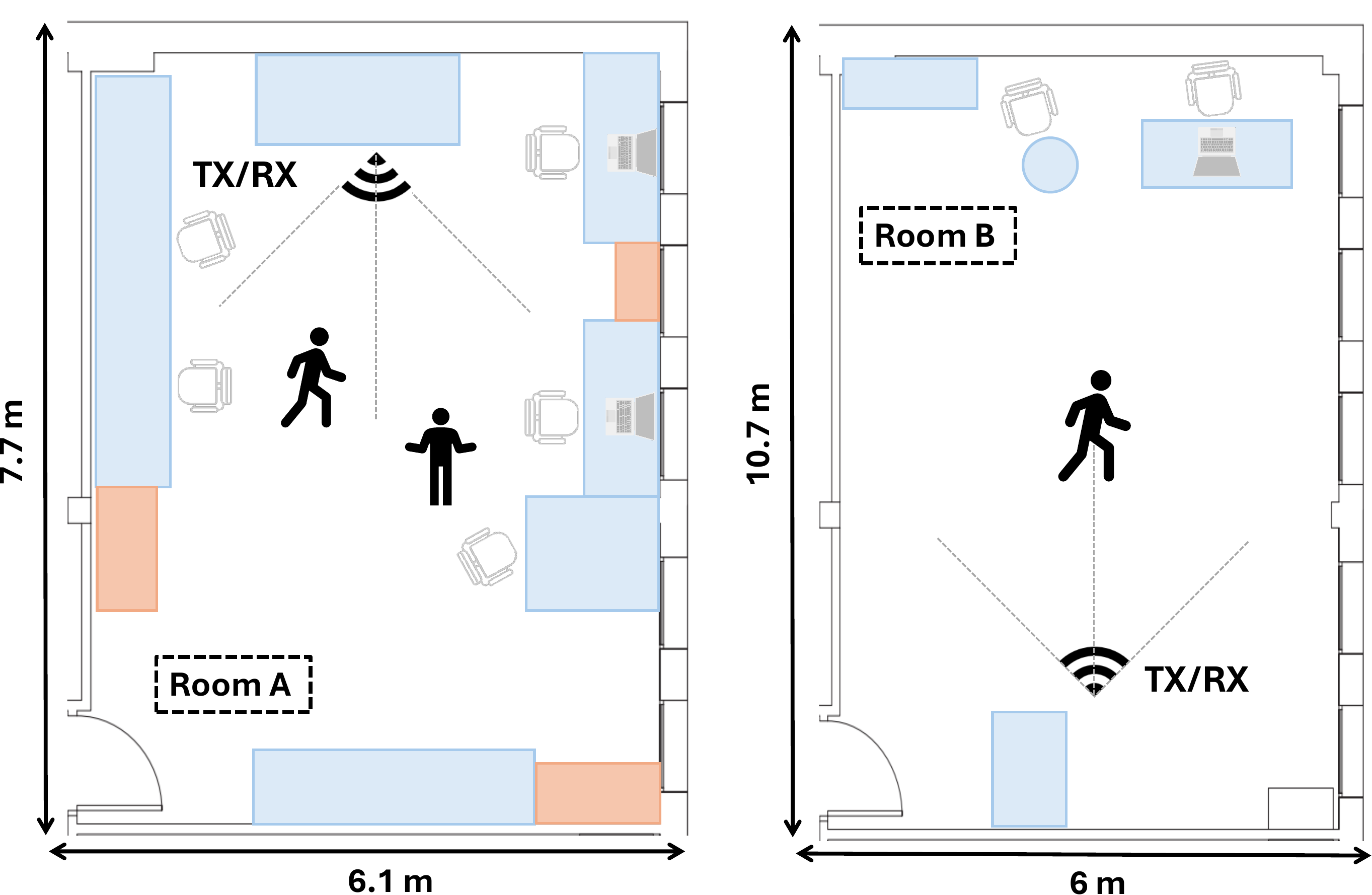}
     \caption{Data collection environments.}
     \label{fig:env}
\end{figure}
The experiments included in DISC-A, most of DISC-B, and DISC-C are performed in a laboratory of $6.1$ $\times$ $7.7$~meters (room A) with a complex multi-path environment due to additional scattering and reflections caused by furniture, computers, screens, and a wide whiteboard, as shown in \fig{fig:env} on the left. The testbed is positioned at one of the two shorter sides of the room, oriented towards the whiteboard. This leaves sufficient space for the subjects to move and perform the different activities included in the dataset. 
On the right of \fig{fig:env} we show the second room used in our dataset (room B), which is a $10.7 \times 6$ meeting room. DISC-B includes \ac{cir} measurements collected in room B to allow testing \ac{isac} algorithm across different environments.

\subsection{Full-duplex ISAC testbed implementation}

\begin{figure}
     \centering
     \includegraphics[width=1\columnwidth]{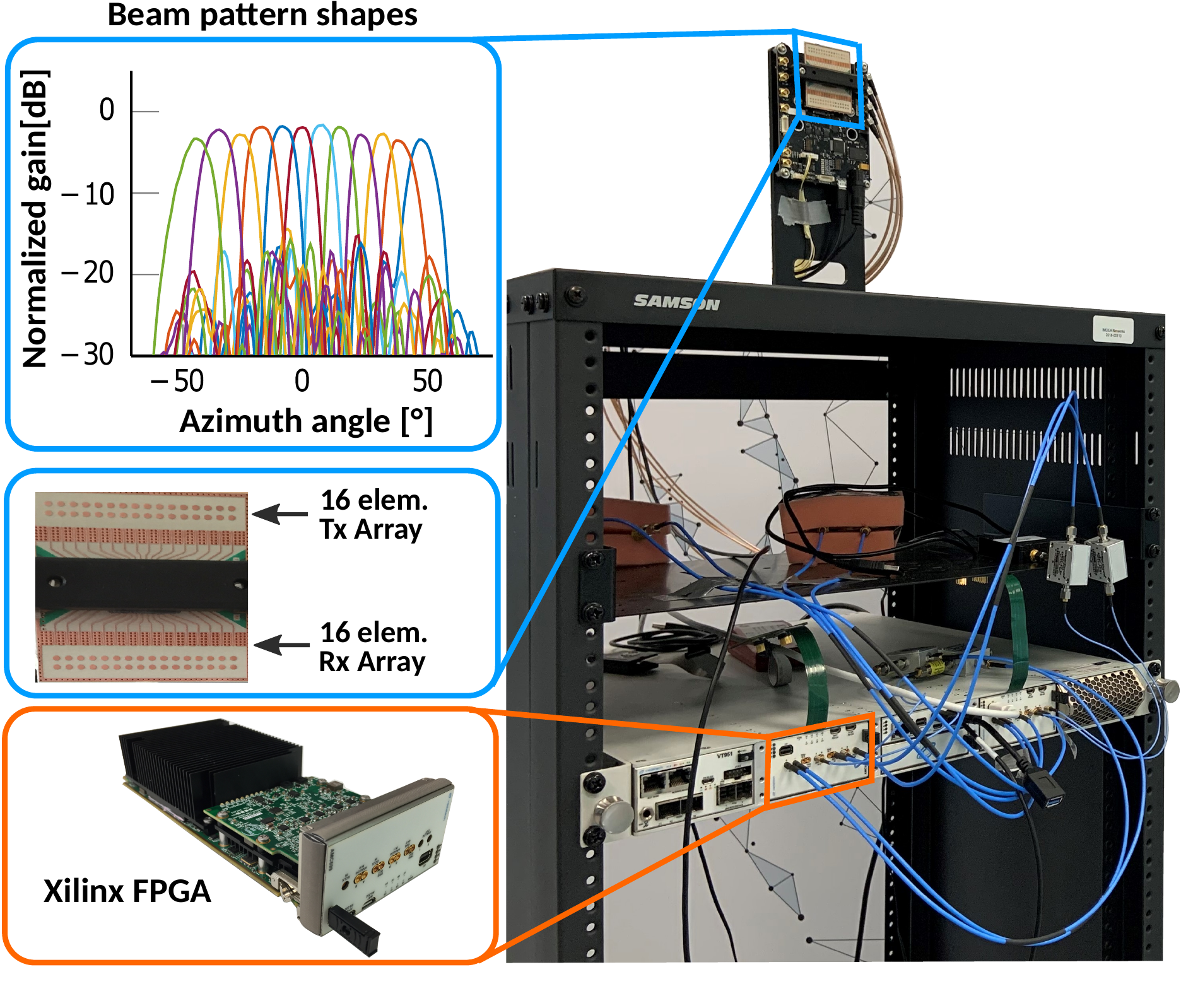}
     \caption{Testbed used in the data collection.}
     \rev{\label{fig:equip}}
\end{figure}

To collect the dataset, we implement a \ac{mmwave} \ac{isac} testbed using the open-source mm-FLEX platform \cite{Lacruz_MOBISYS2020} as a baseline design. 
The baseband processor is based on Vadatech slot cards, integrating a Xilinx UltraScale \ac{fpga}, multi-Giga-sample AD/DA converters, $20$~GB of RAM, and a high-end microprocessor. 
The \ac{mmwave} front-end includes a Sivers EVK$06002$ development kit with $60$~GHz up/down converters to implement IEEE~802.11ad/ay channels. 
It integrates phased arrays including $16$ antenna elements for both the transmitter (TX) and the receiver (RX) \ac{rf} chains. 
These are equipped with phase shifters, enabling analog beamforming. The main components of the testbed are shown in \fig{fig:equip}. 

For the data collection, we configure the Analog-to-Digital/Digital-to-Analog (AD/DA) converters to operate at $3.52$~Gsps sampling frequency. 
We enable fast antenna beam reconfiguration capability during the transmission of a packet to estimate the \acp{cir} using different \ac{bp} shapes in the same packet. 
To provide the required \ac{isac} functionality, we enhance the testbed with new features, including full-duplex capability, synchronization between TX and RX datapaths, and variable \ac{ifs} 
to emulate real traffic patterns.  


We modified the testbed to support \ac{cir} data collection with variable \ac{ifs}, supporting Wi-Fi traffic patterns. Further details regarding this aspect and how to implement full-duplex capabilities are given in~\cite{disc}.

To enable \ac{aoa} estimation techniques \cite{Lacruz_MOBISYS2020}, we measure the \acp{bp} in the codebook of the Sivers kits in a semi-anechoic chamber, with a granularity of $0.5^\circ$~\cite{Lacruz_MOBISYS2020}. For the experiments, we select a subset of \acp{bp} that covers uniformly the \ac{fov} of the antenna, which are shown in \fig{fig:equip}. 

As part of DISC, we provide the bitstream and executable software for the baseband processor, as well as MATLAB$^{\copyright}$ scripts to generate packets and process captured data from the testbed~\cite{disc}.

\section{DISC overview}\label{sec:dataset}

In this section, we provide a high-level overview of the DISC dataset. 
DISC is available at~\cite{disc}, together with extensive documentation, and is complemented by a code repository containing Python code to process the data and replicate our benchmarks in \secref{sec:benchmarks}. Informed consent was obtained from all the subjects involved in the data collection.

DISC consists of three parts, detailed in the following.

\textit{1) DISC-A: Uniform IFS, forward-looking, multiple subjects.} The first part of the dataset, termed DISC-A, contains $433$ \ac{cir} sequences for a total of over $1$ hour of time. Such sequences are collected with uniform \ac{ifs} $T=0.27$~ms and contain the contribution of $7$ different subjects performing $4$ activities: \textit{walking}, \textit{running}, \textit{sitting down/standing up}, and \textit{waving hands}. Each sequence is collected with a single subject present in room A. Channel estimation is performed by appending a single TRN unit to each transmitted packet, using a \ac{bp} pointing forward, along the antenna boresight. 

\textit{2) DISC-B: Uniform IFS, multiple \acp{bp}, multiple subjects moving freely and concurrently.} In the second part of the dataset, we provide over $40$ minutes of \ac{cir} sequences collected with the same uniform \ac{ifs} used in DISC-A. However, in DISC-B we append $12$ TRN fields to each packet, steering the \ac{bp} in each of them to scan the whole field of view of the antenna. This enables the estimation of the \ac{aoa} of the signals, and hence to localize and track subjects in the environment. 

The \ac{cir} sequences in DISC-B are collected across $7$ different days.
A first set of \ac{cir} sequences contains a single subject moving freely in the room and performing one of $5$ possible activities: \textit{walking}, \textit{running}, \textit{sitting down/standing up}, \textit{waving hands}, and \textit{standing still}. A second set of sequences contains multiple subjects \textit{concurrently} present in room A (up to a maximum of $5$ subjects), in different locations, and performing different activities. \rev{This part of the dataset also includes measurements obtained with a \textit{second} ISAC transceiver, located $1.8$~m from the main one. This can be leveraged to explore sensing algorithms that exploit multiple points of view on the environment.} Finally, DISC-B provides data collected in room B, to enable testing \ac{isac} algorithms across different multipath environments. 

\textit{3) DISC-C: Non-uniform IFS, multiple \acp{bp}, a single subject moving freely.} In the third part of DISC, we provide sparse \ac{cir} measurements collected according to the traffic patterns of real Wi-Fi \acp{ap}. 
To collect such measurements, we exploit the configurable \ac{ifs} provided by our testbed and schedule the packet transmissions using the public \texttt{pdx/vwave} dataset  \cite{pdx-vwave-20090704}. This contains real traffic traces captured in different environments using sub-$6$~GHz Wi-Fi \acp{ap}. 
For our experiments, we select representative traces of three types of real-life environments: a computer science department, a library, and an internet cafe. 
In addition to the packet transmissions contained in the \texttt{pdx/vwave} traces, the sparse sequences contain additional transmissions (\textit{injections}) following the protocol described in~\cite{pegoraro2022sparcs}. 

In DISC-C, we collect data in room A, using the same channel estimation procedure used in DISC-B, i.e., transmitting $12$ TRN units with different \acp{bp} for \ac{aoa} estimation. A single subject is present in each \ac{cir} sequence, performing the same $4$ activities of DISC-A.

\rev{Note that, in the dataset collection, each subject was asked to perform one of the pre-defined activities given above. However, each subject performs the activity in a unique way, introducing subject-specific diversity.}
For all the three parts of DISC, we provide (i)~the raw \ac{cir} measurements as multi-dimensional arrays in MAT-file format (\texttt{.mat}) and (ii)~pre-computed \ac{md} spectrograms that can be readily used to develop \ac{ml}-based \ac{har} or person identification algorithms.

\section{Envisioned use cases}\label{sec:discussion}
We identify and discuss five main possible use cases for the dataset.

\subsubsection{People tracking}
Estimating the position of people across time serves as an enabler for a vast number of applications including people flow control, remote healthcare, and intrusion detection, among others. In \ac{isac}, people cause scattering components that are detected as energy peaks in the channel response and can be used to localize and track people across time.
People tracking is particularly challenging in multitarget scenarios, where multiple people are concurrently moving in the same indoor space. This is especially true for \ac{mmwave} signals, which are easily blocked by the human body and are thus prone to mutual occlusion among the subjects. DISC-B contains \ac{cir} sequences collected in such challenging multitarget situations, with up to $5$ subjects.

Moreover, under sparse traffic patterns, people tracking algorithms deserve additional attention as not having regularly sampled measurements can degrade the tracking accuracy. Standard methods based on the \ac{ekf} applied to the range-angle position measurements may need to be adapted~\cite{pegoraro2023rapid}.
DISC-C allows designing and validating new people tracking methods based on \ac{ml} and traditional signal processing. 

\subsubsection{\ac{md} reconstruction} 

Human movement causes a complex superposition of multiple Doppler shifts on the received signal, caused by the different body parts, i.e., a \ac{md} effect. Such \ac{md} can be extracted from a sequence of uniformly spaced \ac{cir} estimates, as shown in \cite{pegoraro2023rapid}, by applying \ac{stft}. The resulting spectrogram serves as a feature of human movement that can be used for downstream tasks such as \ac{har} and person identification.

In the more realistic case where the \ac{cir} estimates are not uniformly spaced, standard \ac{stft} would yield a corrupted \ac{md} spectrogram. Therefore, more advanced \ac{md} extraction should be applied, see for example the one in~\cite{pegoraro2022sparcs}, based on \ac{cs}.
For further details on \ac{md} signatures, including how to extract them from uniformly sampled and sparse measurements, we refer to \cite{pegoraro2023rapid, pegoraro2022sparcs}.

Using DISC-C, which contains \ac{cir} measurements collected with real \ac{ifs} from Wi-Fi traffic, one can develop and test novel sparse reconstruction algorithms to extract \ac{md} signatures of the subjects' movement. As a benchmark solution, we refer to \cite{pegoraro2022sparcs}.
This is an important research direction to enable \ac{isac} in realistic scenarios where the inter-packet duration is dictated by communication rather then by the sensing performance.

\subsubsection{Activity recognition in \ac{isac}}
Our dataset provides the opportunity to validate \ac{har} \ac{ml} algorithms on the \ac{cir} data. 
A \ac{ml} classifier can be trained to distinguish between different activities performed by the subjects, based on suitable features extracted from the \ac{cir} such as the \ac{md} signatures of the movement. A common signal feature that is used for \ac{har} is the \ac{md} spectrogram, which can be computed as detailed in point 2).

\subsubsection{Gait-based identification in \ac{isac}}

By selecting the measurements containing the walking activity from the $7$ subjects in DISC-A, one can train and validate classifiers to perform subject identification based on their gait.
A typical way is to extract the \ac{md} signature of the gait for each person and train a classifier to learn subject-specific patterns that reveal their identity.
This is a challenging task that requires fine-grained feature extraction capabilities and is typically performed with \ac{dl} methods.
Ours is the first dataset to enable such task with IEEE 802.11ay \ac{cir} sequences, which makes it substantially different from existing \ac{mmwave} radar datasets. 

\subsubsection{Domain adaptation to sparse data} A more challenging task entails training the classifiers on the uniformly sampled data, and then testing it on the sparse \ac{cir} measurements. This requires a higher generalization capability as the reconstructed \ac{md} signatures from sparse data have a lower (and varying) resolution. Domain adaptation techniques could be applied to solve this problem.

In the next section, we provide benchmark results for use cases $1)$, $2)$, $3)$, and $5)$, since they represent the most widely studied \ac{isac} applications.

\section{Example results}\label{sec:benchmarks}

    \begin{table}[t!] 
	\caption{Benchmarks for ML models.} \label{tab:benchmarks-results}
    \vspace{-0.25cm}
	\begin{center}
		\begin{tabular}{ccccc}
			\toprule	
            \textbf{Dataset} & \textbf{Activity} & \textbf{F1-score}\\
            \midrule
            \multirow{4}{*}{DISC-A}&Walking
        &0.996\\ 
            &Running&0.974\\ 
            &Sitting/standing&0.947\\ 
            &Waving hands&0.928\\ 
            \midrule
            \multirow{5}{*}{DISC-B}&Walking&0.993\\ 
            &Running&0.976\\ 
            &Waving hands&0.894\\
            &Sitting/standing&0.961\\
            &Standing still&0.926\\
            \midrule
            \multirow{4}{*}{DISC-C}&Walking&0.765\\ 
            &Running&0.742\\ 
            &Sitting/standing&0.773\\
            &Waving hands&0.021\\
			\bottomrule
		\end{tabular} 		
	\end{center}
    
\end{table}

This section presents benchmark algorithms and example results based on DISC. We release their code implementation on GitLab, the link is provided at~\cite{disc}.

\subsection{People tracking}
From the raw IQ samples available in the dataset, it is possible to perform localization and tracking of people in the environment. This is achieved by using the \ac{cir} estimates collected over time, which must be processed in three steps explained in~\cite{pegoraro2023rapid}: 
\textit{(i)} subtraction of the background channel response caused by static objects, \textit{(ii)} estimation of the targets' distances and angles with respect to the \ac{isac} device, and \textit{(iii)} using a tracking algorithm (e.g., a Kalman filter) to estimate each person's movement trajectory across time. 
The localization step (ii) is carried out using the magnitude peaks of the \ac{cir} estimates, which correspond to \textit{observations} of the positions of the subjects in the surroundings.
In step (iii), in our results we employ an \ac{ekf} to track the physical position of each individual in the Cartesian space from the localization measurements. A constant velocity model is used to approximate the movement of the subject.
The association between the observations from subsequent timesteps is done using the Nearest-Neighbors Joint Probabilistic Data Association algorithm (NN-JPDA)~\cite{shalom2009probabilistic}. 

\begin{figure}[t!]
	\begin{center}   
		\centering
		\subcaptionbox{Trajectories of $5$ subjects from DISC-B. Each subject performs an activity among the $5$ present in the dataset. Grey dots represent measurements associated with tracks. \label{fig:track}}[4.1cm]{\includegraphics[width=4.1cm]{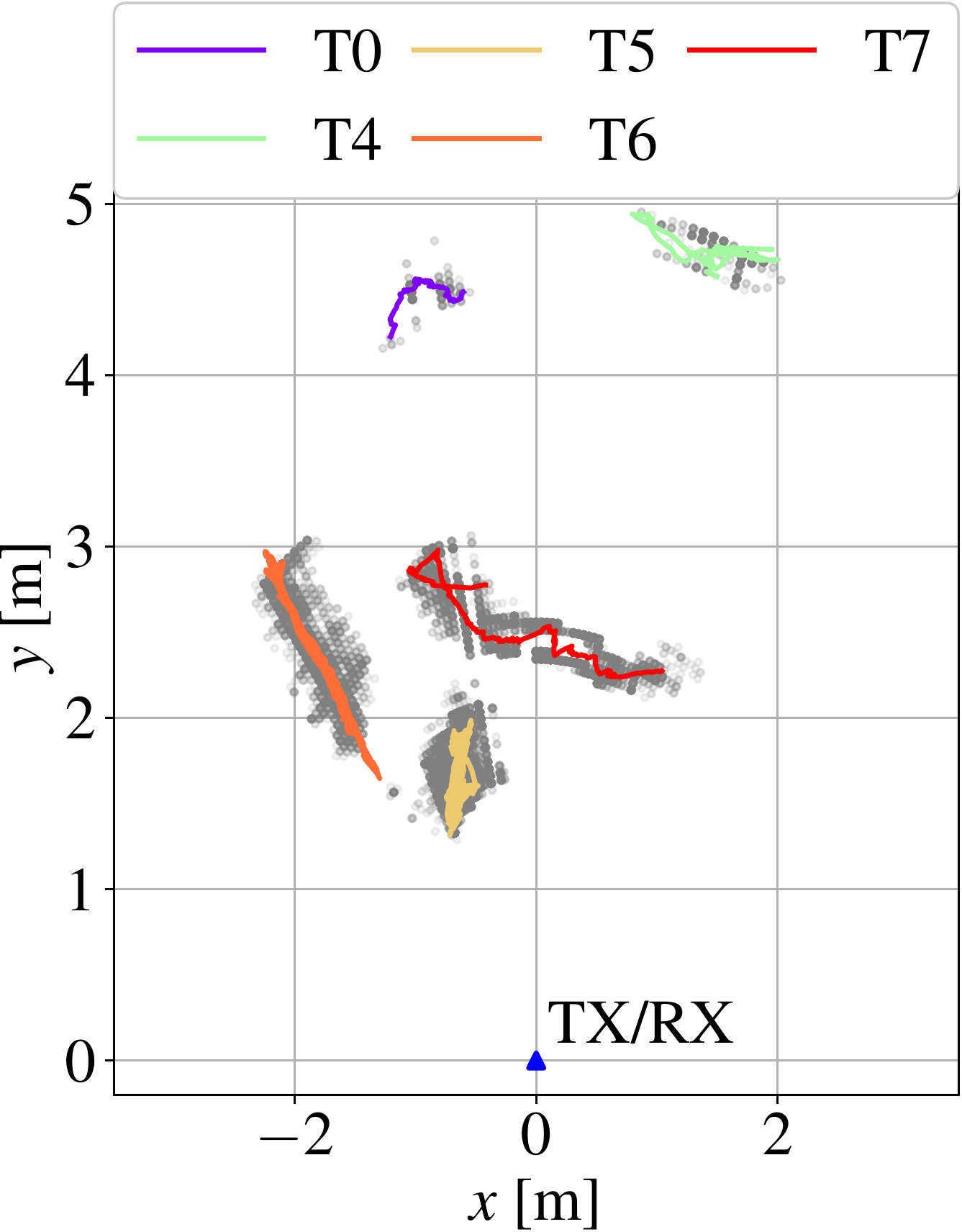}}
		\subcaptionbox{\ac{md} spectrograms corresponding to T$4$, T$5$, ad T$6$, corresponding to activities \textit{waving hands}, \textit{sitting/standing}, and \textit{running}. Each activity presents a unique pattern.  \label{fig:md}}[4.4cm]{\includegraphics[width=4.4cm]{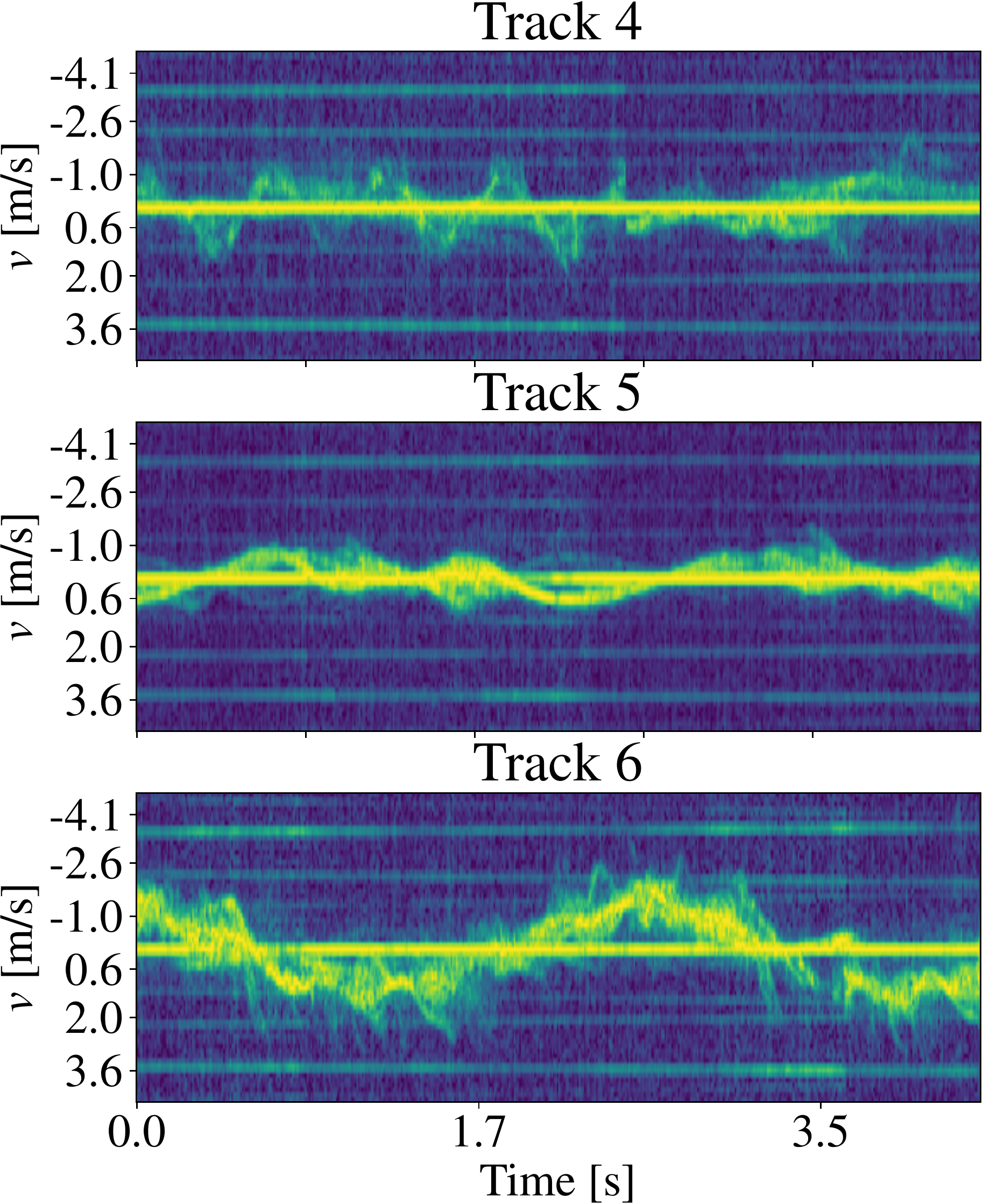}}
		\caption{Example tracks obtained from a \ac{cir} sequence in DISC-B (a) and the corresponding \ac{md} spectrograms (b).}
		\label{fig:track-multitarget}
	\end{center}
    \vspace{-3mm}
\end{figure}

The results of this process are shown in \fig{fig:track}, in which $5$ subjects are tracked obtaining $5$ tracks (denoted by T$x$ where $x$ is an identifier of the track) from an example \ac{cir} trace from \mbox{DISC-B}. Each subject is performing an activity: T$0$ \textit{sitting/standing}, T$4$ \textit{waving hands}, T$5$ \textit{sitting/standing}, T$6$ \textit{running}, and T$7$ \textit{walking}. In the next section, we show that DISC allows extracting \ac{md} sequences from each track to recognize the activities performed by the subjects.

\vspace{2mm}
\subsection{\ac{md} extraction}

\rev{To provide a benchmark algorithm to extract \ac{md} spectrograms of human movement, we apply \ac{stft} to the \ac{cir} sequences,  as described in \secref{sec:11ay}, after localizing the target and extracting the \ac{cir} part corresponding to the target's location. For the \ac{stft} we use a Hamming window of $W=64$ samples and normalize the resulting \ac{md} spectrogram column-wise in $[0, 1]$, using $\min-\max$ normalization.}
Thanks to the \ac{md} signatures, in \fig{fig:md} it is possible to differentiate between the different activities performed by each subject. We report the activities corresponding to T$4$-$6$, throughout $4$~s. The resulting \ac{md} pattern is typical of each activity. As an example, for \textit{running} (T$6$) it shows the strong contribution of the torso and the fainter one of the moving limbs.

\begin{figure}[t!]
	\begin{center}   
		\centering
		\subcaptionbox{ \label{fig:corrupted}}[4.3cm]{\includegraphics[width=4.3cm]{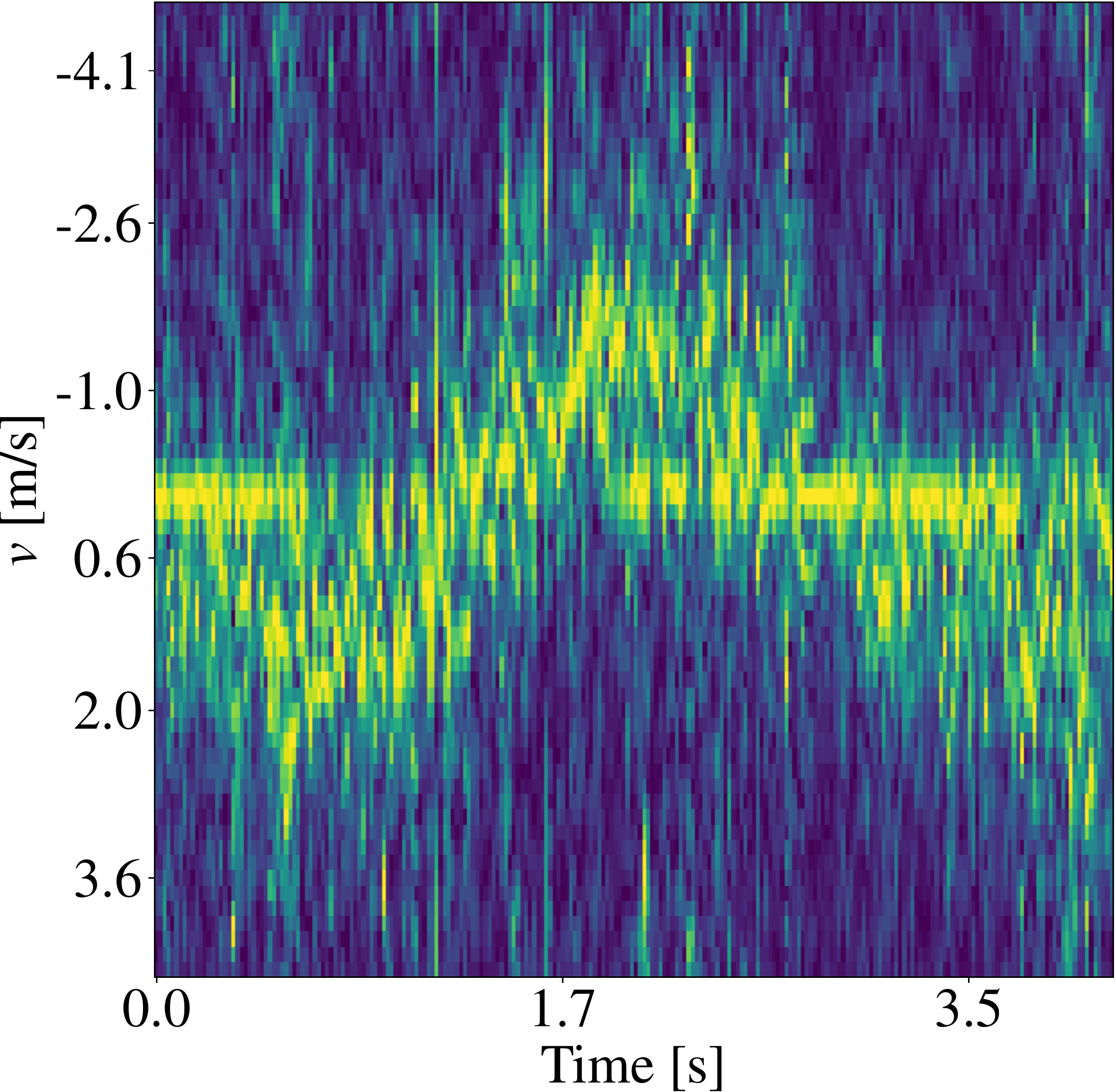}}
		\subcaptionbox{\label{fig:sparse}}[4.3cm]{\includegraphics[width=4.3cm]{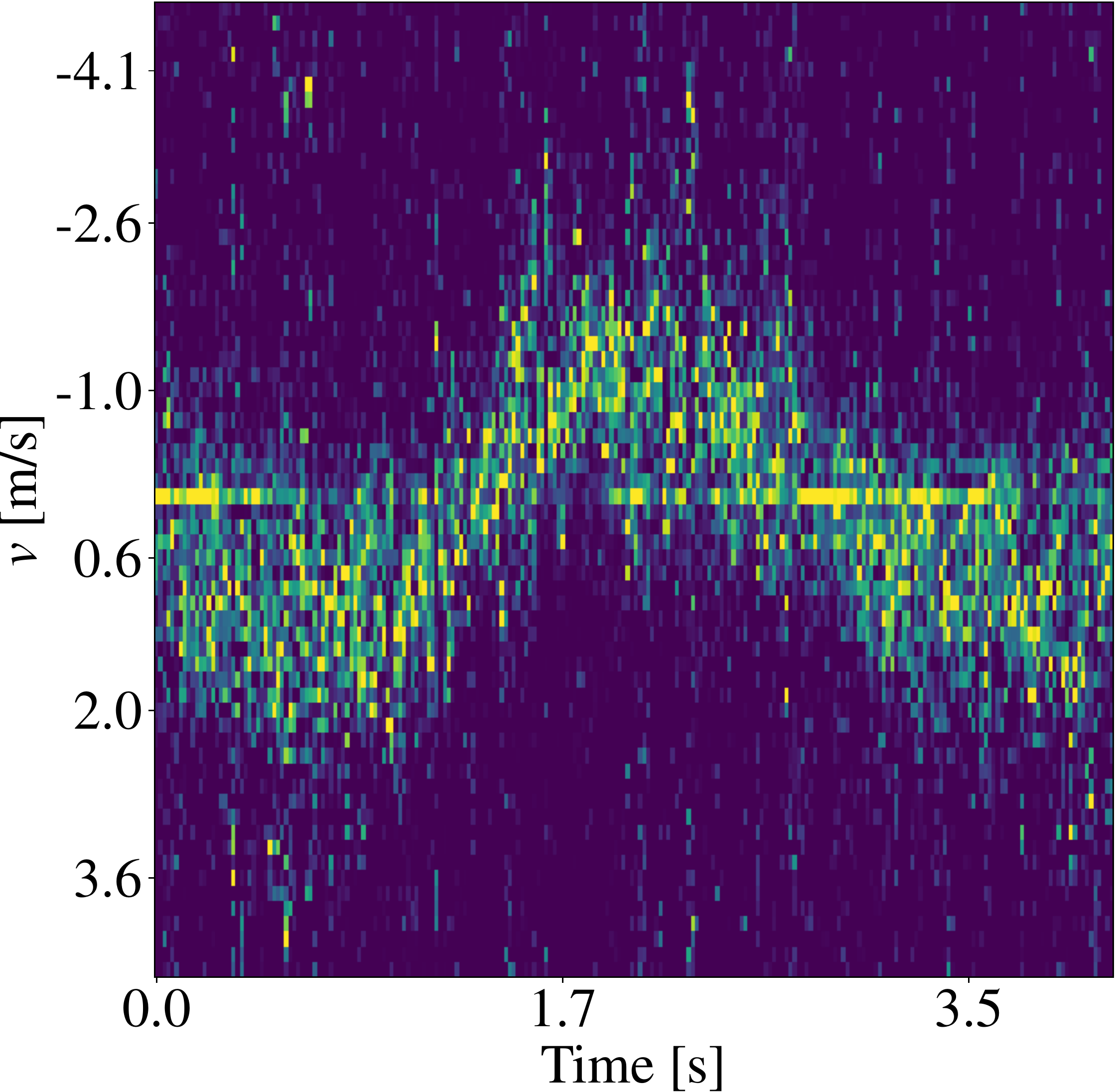}}
		\caption{\ac{md} spectrogram of a walking subject from irregular \ac{cir} measurements from DISC-C: (a)~using \ac{stft} and (b)~\ac{iht}~\cite{pegoraro2022sparcs}.}
		\label{fig:sparse-md}
	\end{center}
    \vspace{-3mm}
\end{figure}

As part of \mbox{DISC-C}, we also release \ac{md} spectrograms obtained from non-uniformly sampled \ac{cir} sequences, using the \ac{iht} algorithm as detailed in~\cite{pegoraro2022sparcs}, and the corresponding code implementation. We report an example of spectrograms reconstructed with \ac{iht} in \fig{fig:sparse-md}, in comparison with those obtained using standard \ac{stft} on irregularly sampled \ac{cir} data. 

\subsection{Human activity recognition}
Different activities performed by subjects cause different, time-varying patterns in the Doppler shift of the received signal. These are contained in the phase of the \ac{cir} data, and are, in turn, reflected in the \ac{md} signatures. Therefore, it is possible to train \ac{dl} models for multi-class classification problems that receive as input the \ac{md} signature and output the most likely activity performed by the subject. In the following, we present: 1) a basic benchmark to demonstrate the feasibility of \ac{md}-based activity recognition from the DISC dataset, and 2) a more advanced task in which the classifier is trained on uniformly sampled \ac{cir} data and tested on irregular and sparse communication traffic.  

\subsubsection{Basic classification benchmark} To provide a benchmark \ac{dl} classifier, we developed a basic \ac{cnn} model, composed of (i)~six convolutional layers for feature extraction, with $3\times 3$ kernel size, a stride of $2$, and $8, 16, 32, 64, 128, 128$ feature maps, and (ii)~a classifier with two fully connected layers with $64$ and $C$ neurons, respectively, where $C$ is the number of activities in the \ac{har} task. Dropout with probability $0.2$ is applied before each fully connected layer during training.

We trained the model for $10$ epochs on the \ac{md} signatures from DISC-A and DISC-B (see \secref{sec:dataset}) independently. 
In \tab{tab:benchmarks-results}, we show the classification \textit{F1 score}, defined as the harmonic mean of precision and recall, on the test set. In DISC-A, we include \ac{cir} sequences from subjects $1$ to $5$ in the training set, while subjects $6$ and $7$ are used for the validation and test sets, respectively. This is intended to demonstrate the possibility of generalizing to unseen subjects, thus testing the generalization capabilities of the \ac{dl} classifier.
In DISC-B, the validation and test sets are a randomly selected $10$\% fraction of the total dataset each.

\rev{Our results demonstrate the feasibility of performing \ac{har} with radar-like accuracy using standard-compliant \ac{sc} IEEE~802.11ay waveforms.} From \tab{tab:benchmarks-results}, one can see that the most challenging activity to recognize in both datasets is \textit{waving hands}. Indeed, this activity only involves the movement of arms, which do not cause strong scattering of the radio signal and are hence difficult to detect.

\subsubsection{Generalization to sparse and irregular \ac{cir} data}
The temporal irregularity and sparsity of \ac{cir} measurements obtained from normal communication traffic degrade the quality of the resulting \ac{md} spectrograms.
Moreover, collecting data containing diverse temporal \ac{cir} estimation patterns is challenging and time-consuming. However, classifiers trained on \ac{cir} data collected at regular sampling times may not generalize well to irregular patterns, due to the reduced \ac{md} quality.
To show this phenomenon, we train a \ac{cnn} classifier like the one used in the previous section on \mbox{DISC-B}. Then, we test it on data contained in \mbox{DISC-C}, evaluating the generality of the learned features. 

The results are shown in \tab{tab:benchmarks-results}. The sparsity of the data degrades the performance of the model in three of the activities (\textit{walking}, \textit{running}, and \textit{sitting/standing}), which are however still recognizable with over $0.7$ F$1$-score. For the last activity (\textit{waving hands}), the model does not generalize enough to the new data. This activity is more challenging than the other three since it involves movements of the hands that cause weak signal scattering. This shows the importance of addressing the problem of \textit{domain shift} and generalization in \ac{dl}-based \ac{isac}. 

\rev{\section{Limitations and future work}\label{sec:limitations}
While \mbox{DISC} serves as a valuable benchmark dataset for \ac{isac} research, certain limitations present opportunities for further extensions and improvement. 

\textit{Dataset size.} DISC contains a sufficient amount of data to train and validate \ac{dl} models for \ac{isac}. However, compared to existing \ac{dl} datasets in other fields (e.g., computer vision), its size is fairly limited in terms of total measurement time, which may limit the applicability of extremely large \ac{dl} architectures. 

\textit{Occlusion and interference.} Although \mbox{DISC} captures two distinct indoor environments, five different activities, and up to five subjects moving simultaneously, acquiring data in more diverse settings, including the presence of obstacles in future work would enhance the value of the dataset.

\textit{Bistatic and multistatic CIR.} As research moves towards \textit{cooperative} \ac{isac} networks, it is key to collect datasets including bistatic and multistatic \ac{cir} data, with multiple distributed \ac{isac} TX and RX nodes. 
This is a critical aspect to be investigated by future work and \ac{isac} datasets.

\textit{Irregular temporal patterns generalization.} While \mbox{DISC-C} includes realistic Wi-Fi traffic patterns, its irregular and sparse nature introduces difficulties in \ac{md} extraction and classification, requiring adaptive signal processing techniques and domain adaptation strategies for \ac{dl} architectures. 
Future research should focus on developing and evaluating algorithms that generalize across different inter-packet durations while maintaining robust sensing performance.


}

\section{Conclusion}\label{sec:conclusion}
In this paper, we presented DISC, the first \ac{isac} dataset containing \ac{cir} sequences obtained using $60$~GHz, standard-compliant, IEEE~802.11ay waveforms that contain backscattered signals on people moving in the environment. The provided \ac{cir} measurements include $7$ subjects performing $5$ different activities in $2$ environments, and they are collected according to both uniform and real Wi-Fi traffic patterns, thus enabling the validation of a variety of \ac{ml} and signal processing algorithms for \ac{isac}.
We presented benchmark algorithms for people tracking, \ac{md} extraction, and activity recognition under diverse channel estimation patterns, and made them available to foster further \ac{isac} research.


\ifCLASSOPTIONcaptionsoff
  \newpage
\fi

\bibliography{references}
\bibliographystyle{ieeetr}

\renewenvironment{IEEEbiography}[1]
  {\IEEEbiographynophoto{#1}}
  {\endIEEEbiographynophoto}

	\begin{IEEEbiographynophoto}
{Jacopo Pegoraro} (M'23) is an Assistant Professor in the Department of Information Engineering (DEI) at the University of Padova, Italy.
\end{IEEEbiographynophoto}
\vskip 0pt plus -1fil
\begin{IEEEbiographynophoto}{Pablo Saucedo}
is a PhD student at IMDEA Networks in Madrid, Spain.
\end{IEEEbiographynophoto}
\vskip 0pt plus -1fil
\begin{IEEEbiographynophoto}{Jesus O. Lacruz}
is a Senior Researcher at IMDEA Networks in Madrid, Spain, since 2017.
\end{IEEEbiographynophoto}
\vskip 0pt plus -1fil
\begin{IEEEbiographynophoto}
    {Michele Rossi} (SM'13) is a Full Professor at the Department of Information Engineering and Department of Mathematics of the University of Padova, Italy. 
\end{IEEEbiographynophoto}
\vskip 0pt plus -1fil
\begin{IEEEbiographynophoto}{Joerg Widmer}
(
F'20) 
is Research Professor and Research Director of IMDEA Networks in Madrid, Spain. 
\end{IEEEbiographynophoto}

\end{document}